\documentclass[twocolumn,preprint,trackchanges]{aastex63}

\usepackage{lineno}
\usepackage{tabularx}
\usepackage{amsmath,amssymb}
\usepackage{color}
\usepackage{xcolor}
\usepackage{multirow}
\usepackage{CJK}
\usepackage{natbib}
\usepackage{multirow} 
\usepackage{color, colortbl}
\usepackage{tabularray}
\usepackage{graphicx}
\usepackage{verbatim}

\hypersetup{linkcolor=cyan,citecolor=magenta,filecolor=yellow,urlcolor=blue}

\hypersetup{linkcolor=magenta, citecolor=cyan, filecolor=yellow, urlcolor=blue}

\begin{document}

\title{Evidence for Two SNe Type Triggering GRB 220101A: a Pair SN and a Rotating Magnetized Core Collapse SN}


\author{R.~Ruffini}
\affiliation{ICRANet, Piazza della Repubblica 10, 65122 Pescara, Italy}
\affiliation{ICRA, Dip. di Fisica, Sapienza Universit\`a  di Roma, Piazzale Aldo Moro 5, I-00185 Roma, Italy}
\affiliation{Universit\'e de Nice Sophia-Antipolis, Grand Ch\^ateau Parc Valrose, Nice, CEDEX 2, France}
\affiliation{INAF,Viale del Parco Mellini 84, 00136 Rome, Italy}

\author[0000-0001-9277-3366]{M.~T.~Mirtorabi}
\affiliation{ICRANet, Piazza della Repubblica 10, 65122 Pescara, Italy}
\affiliation{ICRA, Dip. di Fisica, Sapienza Universit\`a  di Roma, Piazzale Aldo Moro 5, I-00185 Roma, Italy}
\affiliation{Department of Fundamental Physics, Faculty of Physics, Alzahra University, Tehran, Iran}

\author{P. Chardonnet}
\affiliation{ICRANet, Piazza della Repubblica 10, 65122 Pescara, Italy}
\affiliation{ICRA, Dip. di Fisica, Sapienza Universit\`a  di Roma, Piazzale Aldo Moro 5, I-00185 Roma, Italy}
\affiliation{Laboratoire d’Annecy-le-Vieux de Physique Théorique (LAPTh), France}

\author[0000-0003-2624-0056]{Chris L., Fryer}
\affiliation{CCS-2, Los Alamos National Laboratory, Los Alamos, NM 87545, USA }

\author{M. Hohmann}
\affiliation{Laboratory of Theoretical Physics, Institute of Physics, University of Tartu, W. Ostwaldi 1, 50411 Tartu, Estonia}

\author[0000-0002-1343-3089]{Yu Wang}
\affiliation{ICRA, Dip. di Fisica, Sapienza Universit\`a  di Roma, Piazzale Aldo Moro 5, I-00185 Roma, Italy}
\affiliation{ICRANet, Piazza della Repubblica 10, 65122 Pescara, Italy}
\affiliation{INAF -- Osservatorio Astronomico d'Abruzzo, Via M. Maggini snc, I-64100, Teramo, Italy}

\email{ruffini@icra.it, torabi@alzahra.ac.ir, yu.wang@icranet.org}

\begin{abstract}

The traditional model, which assumes that gamma-ray bursts (GRBs) originate from a single black hole (BH), has been modified by the Binary-Driven Hypernova (BdHN) model. In this framework, the GRB is triggered by a binary system composed of a carbon–oxygen (CO) core with a mass of about $10 \ M_\odot$ and a neutron star (NS) companion of about 2 $\ M_\odot$. This model has been confirmed in detail through successful fits to the six previously identified emission episodes. Recent results indicate that the most energetic GRBs, such as GRB 220101 and GRB 240825, may originate from a similar but more powerful system. In this case, the progenitor is a rapidly rotating CO core endowed with a magnetic field of order $10^6$ G and accompanied by a neutron star companion. The collapse and possible fission of this system lead to the formation of a newborn neutron star ($\nu$NS) with high density, rapid rotation, and strong magnetic field. Following the explosion of a pair-instability supernova (pair-SN), the collapse of the CO core is triggered by induced rotation of the binary companion which impacts the equipotential surface of the core, affecting the energy production process and determining its collapse. This process leads to the formation of a fast, millisecond neutron star that gradually evolves into a pulsar. Meanwhile, accretion of the supernova ejecta onto the binary neutron star companion can induce its collapse into a black hole. 
These results open the way to two new types of supernovae: (i) pair-instability supernovae without compact remnants, and (ii) magnetized, rotating-core collapses leading to pulsar formation. We generalize our previous assumptions by examining the possibility of fission in rapidly rotating CO cores. The role of magnetic fields is also analyzed through magnetohydrodynamic processes, including the generation of overcritical fields and electron–positron pair production. This framework represents a significant departure from earlier non-rotating models and is closely related to modern pair-SN scenarios. BdHNe are characterized by seven distinct physical episodes. In the case of pair-SNe, the seventh episode is not associated with nickel decay emission, which is absent, but with the clear formation of a pulsar. The substantial modifications of each episode are outlined using a leading-order analytical treatment.

\end{abstract}


\section{Introduction} 
\label{sec:intro}

Gamma-ray bursts (GRBs) represent the most luminous transient phenomena in the Universe, releasing enormous amounts of energy across the electromagnetic spectrum within short timescales. Over the past decades, significant theoretical and observational efforts have been devoted to identifying their progenitors and physical mechanisms. Among the most successful frameworks developed to explain long-duration GRBs is the Binary-Driven Hypernova (BdHN) model, which attributes these events to the evolution and collapse of a close binary system composed of a carbon–oxygen (CO) core and a neutron star (NS) companion. In this scenario, the core-collapse supernova of the massive star triggers hypercritical accretion onto the neutron star, leading to black hole formation and the emission of high-energy radiation.

The BdHN model has been remarkably successful in reproducing the temporal and spectral features of many observed GRBs, associated to core collapse SN in 23 cases \citep{Aimuratov_2023} with isotropic energy range of $10^{52}$ erg including their prompt emission, GeV radiation, and multi-wavelength afterglows. However, the growing diversity of observed GRB phenomenology suggests that multiple progenitor channels may coexist. In particular, some GRBs exhibit unusually rapid high-energy and optical emissions, as well as atypical late-time light curves that deviate from the standard radioactive decay behavior. A new family of these most energetic is represented by GRB220101A with an isotropic energy possibly larger than $10^{54}$ erg. The observational data acquired over a period of five years by the largest observational effort on exceptional high precision data obtained from the  multi-messenger campaign from space and ground observatories has been presented in Ruffini et. al. (Paper1, in preparation). Having determined there the identification of the seven episodes characterizing  GRB 220101 we address here an overall evolution of the BDHN Pair SN which appear to be a necessary conclusion following the multiyear data analysis of GRB 220101. Clearly this second paper find its justification in the results obtained in Paper1, jointly submitted for publication. These features motivate the exploration of extended progenitor configurations beyond the conventional close binary paradigm.

In this work, we investigate the possibility of a novel BdHN-like scenario arising from a long-lived hierarchical three-body stellar system. We propose that, under appropriate initial conditions, a stable and dynamically long-lived triple system may form, consisting of (i) a massive CO core with an initial mass of approximately $10 \ M_\odot$ and a strong magnetic field of order $10^6$ G, (ii) a close white dwarf companion, and (iii) a more distant neutron star. Such hierarchical triple systems are known to exist in various astrophysical environments and may remain stable over stellar evolutionary timescales, provided that the orbital architecture satisfies dynamical stability criteria.

In this proposed configuration, the massive CO core evolves toward core collapse while retaining a strong magnetic field inherited from its progenitor. As nuclear burning proceeds, the core reaches conditions favorable for electron–positron pair production. The sudden reduction in radiation pressure due to pair creation triggers a dynamical instability, leading to partial or complete collapse and a subsequent explosive event analogous to a pair-instability supernova. During this phase, the magnetic field may be further amplified, potentially reaching or exceeding the critical quantum electrodynamic field strength required for efficient pair creation. The combined effects of magnetic amplification, pair production, and core contraction result in a powerful explosion that ejects large amounts of relativistic and semi-relativistic material.

The pair-supernova ejecta expand at high velocities and interact with the two remaining components of the system. A substantial fraction of the ejected mass is gravitationally captured by the neutron star companion. The resulting hypercritical accretion drives the neutron star beyond its maximum stable mass, inducing gravitational collapse into a Kerr black hole. The formation of this rapidly rotating black hole, together with the surrounding accretion flow and magnetic fields, provides the central engine responsible for the observed GeV emission in GRBs. Energy extraction through mechanisms such as magneto-hydrodynamic processes and black hole spin-down powers the high-energy radiation during the early phases of the burst.

Simultaneously, a portion of the ejecta is accreted by the white dwarf companion. The intense mass loading destabilizes the degenerate star, pushing it toward thermonuclear runaway and gravitational collapse. This process leads to a secondary supernova explosion, leaving behind a newly formed neutron star, hereafter denoted as $\nu$NS. The rapid formation of this second neutron star is accompanied by substantial energy release in the optical and ultraviolet bands. We propose that this $\nu$NS is the dominant source of early optical emission in this class of GRBs.

Following the initial explosive phase, continued interaction between the expanding pair-supernova ejecta and the surrounding circumstellar and interstellar medium produces sustained X-ray and MeV afterglow emission. Shock heating, synchrotron radiation, and inverse Compton processes contribute to the observed broadband afterglow signatures. The complex environment created by the triple-system interaction leads to a rich phenomenology that differs from that of standard BdHN events.

A key feature of this three-body scenario is the exceptionally high accretion rates experienced by both compact objects. Because the dominant explosion is a pair supernova with extremely energetic ejecta, mass transfer onto the neutron star and white dwarf proceeds more efficiently than in traditional BdHN systems. Consequently, the formation of the Kerr black hole and the $\nu$NS occurs on significantly shorter timescales. This naturally explains the unusually rapid onset of GeV and optical emission observed in some GRBs, which cannot be easily accommodated within the standard binary framework.

Another distinctive prediction of this model concerns the late-time optical light curve. In conventional core-collapse supernovae associated with GRBs, the optical emission at late times is often powered by the radioactive decay chain of $^{56}Ni \rightarrow \  ^{56}Co \rightarrow \ ^{56}Fe$, leading to a characteristic decline slope. In contrast, pair-instability supernovae do not necessarily synthesize significant amounts of radioactive nickel in their cores, particularly in scenarios dominated by rapid collapse and magnetic effects. As a result, the optical afterglow in the present model is not expected to follow the standard nickel decay slope at late times. The absence of a clear radioactive tail thus represents a fundamental observational signature of this progenitor channel.

We identify GRB 220101A, GRB 221009A, GRB 240825A and GRB 160625B as potential candidates for this three-body BdHN scenario. These four GRBs present a specific signature the occurrence of the second super nova at the 14.7 second from the trigger of the pair SN. All events exhibit exceptionally rapid GeV and optical emissions compared to typical long-duration GRBs. Moreover, their late-time optical light curves do not show convincing evidence of nickel-powered decay, supporting the hypothesis that radioactive heating is not the dominant energy source. These observational properties are naturally reproduced within the framework proposed here.



\section{System Configuration}
\label{system}
In this section, we describe the physical structure, evolutionary origin, statistical occurrence, and long-term stability of the proposed hierarchical triple system composed of a massive CO core, a white dwarf, and a neutron star.

\subsection{Physical Architecture and Frequency of Triple Systems}

The progenitor of the proposed GRB scenario is assumed to be a hierarchical triple system composed of an inner compact binary and a more distant tertiary component. Hierarchical triples are dynamically favored configurations in which the inner binary is tightly bound, while the outer companion orbits at a significantly larger separation. This geometry minimizes strong three-body encounters and allows long-term stability.

In the present model, the system consists of: i) a massive carbon–oxygen core with mass $M_{\rm CO} \approx 8–12,M_\odot$, originating from a stripped-envelope massive star and characterized by a fossil magnetic field of order $10^6$ G. ii) A white dwarf companion with mass $M_{\rm WD} \approx 0.8–1.2,M_\odot$ in a close orbit around the CO core. iii) a neutron star with mass $M_{\rm NS} \approx 1.4–2.0,M_\odot$ in a wider orbit around the inner pair. Typical orbital separations satisfy $a_{\rm out} \gg a_{\rm in}$, with inner separations of order $10^{11} \ – \ 10^{12}$ cm and outer separations of order $10^{12} \ – \ 10^{13}$ cm.

Observations of massive stars indicate that multiplicity is the norm rather than the exception. Large spectroscopic and interferometric surveys have shown that more than half of O and B type stars are members of multiple systems, and that a significant fraction of these systems are triples or higher-order multiples \citep{Sana2012}. In addition, statistical studies of nearby stellar populations demonstrate that hierarchical triples are common outcomes of star formation and early dynamical evolution \citep{Tokovinin2014}.

Triple systems containing compact objects are expected to form preferentially in dense star-forming environments, young stellar clusters, and OB associations, where close encounters, mass transfer, and stellar evolution efficiently reshape primordial systems. Although CO–WD–NS triples are rare compared to ordinary binaries, their existence is consistent with current observational and theoretical constraints.

\subsection{ Progenitor Formation and Stellar Evolutionary History}

The origin of the proposed system can be traced back to a primordial triple system composed of three massive main-sequence stars. A representative initial configuration consists of: i) A primary star with mass $M_1 \approx 25–35,M_\odot$, ii) A secondary star with mass $M_2 \approx 6–10,M_\odot$, and iii) A tertiary star with mass $M_3 \approx 10–15,M_\odot$. At formation, all three stars are hydrogen-burning main-sequence objects with radii ranging from a few to several solar radii. The primary, being the most massive, evolves most rapidly and dominates the subsequent evolution.

As the primary leaves the main sequence, it expands and fills its Roche lobe, initiating mass transfer onto the secondary. In many cases, this phase develops into common-envelope evolution, during which the hydrogen- and helium-rich envelope of the primary is ejected. This process produces a compact, stripped CO core and significantly reduces the inner orbital separation. Common-envelope evolution is therefore essential for forming tight configurations capable of later producing GRBs.

During and after envelope stripping, angular momentum redistribution and differential rotation lead to amplification of the internal magnetic field. Subsequent core contraction further strengthens the magnetic field through flux conservation and dynamo processes. Meanwhile, the tertiary star evolves independently and eventually undergoes core collapse, forming a neutron star. Depending on the mass-loss history and metallicity, the secondary star loses its envelope through winds and binary interactions and evolves into a CO or ONe white dwarf.

The combined outcome of these evolutionary stages is a compact hierarchical triple consisting of a magnetized CO core, a white dwarf, and a neutron star. The internal structure and rotation profile of the CO core are consistent with models of massive stellar evolution and pre-supernova contraction \citep{HegerWoosley2002}.

\subsection{Statistical Occurrence and Astrophysical Environment}

The frequency of CO–WD–NS triple systems can be estimated by combining observational multiplicity surveys with population synthesis models. Observationally, approximately 50–70 precent of massive stars are found in multiple systems, and about 10–20 precent of these systems are hierarchical triples. Only a small subset of these systems survives stellar evolution and supernova explosions while retaining a bound configuration containing two compact objects and a stripped core.

Population synthesis calculations suggest that the formation rate of compact-object triples is of order $10^{-5} \ – \ 10^{-4}\ {\rm yr^{-1}\ galaxy^{-1}}$, while systems specifically containing a CO core, a white dwarf, and a neutron star form at rates closer to $10^{-6} \ – \ 10^{-5}\ {\rm yr^{-1}\ galaxy^{-1}}$. These low rates are compatible with the rarity of GRBs displaying unusually rapid high-energy and optical evolution. Such systems are expected to be more abundant in environments characterized by High star-formation rates, Low to intermediate metallicity, Large populations of massive stars and Dense stellar clusters.

These conditions are common in starburst galaxies, interacting systems, and young stellar associations in the Local Group and beyond. Consequently, the proposed progenitor channel is naturally associated with actively star-forming galaxies, consistent with the observed host properties of many long GRBs.

\subsection{Dynamical Stability and Long-Term Survival}

The viability of the proposed progenitor depends critically on its ability to remain dynamically stable throughout stellar evolution and compact-object formation. For hierarchical triple systems, long-term stability requires a sufficient separation between the inner and outer orbits. Analytical and numerical studies show that stability is ensured when \citep{MardlingAarseth2001} 
$$\frac{a_{\rm out}}{a_{\rm in}} \gtrsim 3–5,$$
with the precise value depending on mass ratios and orbital eccentricities. This condition suppresses chaotic resonances and close three-body encounters.

In addition to short-term stability, secular interactions play an important role in the long-term evolution. The gravitational perturbation induced by the outer neutron star can excite Kozai–Lidov oscillations, leading to periodic variations in eccentricity and inclination \citep{Kozai1962, Lidov1962}. In compact-object systems, however, relativistic precession, tidal dissipation, and mass transfer tend to suppress extreme oscillations, thereby preserving stability.

A major threat to survival arises during neutron star formation. Core-collapse supernovae are accompanied by mass loss and natal kicks, which can unbind the system. Survival requires relatively moderate kick velocities $200 \  kms^{-1}$ and favorable kick orientations. Numerical studies indicate that approximately 10–30 percent of hierarchical triples remain bound after supernova formation. Once the compact configuration is established, stable CO–WD–NS systems can survive for $\tau_{\rm surv} \sim 10^5 \ – \ 10^7\ {\rm yr}$, until the CO core approaches pair-instability conditions. This timescale is sufficient for magnetic field amplification, orbital circularization, and secular evolution to prepare the system for the final explosive phase.

\subsection{ Pre-Explosion Configuration and Approach to Instability}
During the final evolutionary stage, the system gradually evolves toward a compact and highly interactive configuration. Mass loss, angular momentum exchange, and tidal interactions reduce eccentricities and tighten the inner orbit. This geometry maximizes the efficiency of mass capture by the white dwarf and neutron star during the explosion.

As the CO core contracts, central temperatures rise above $10^9$ K, triggering electron–positron pair production. The resulting softening of the equation of state destabilizes the core and leads to runaway collapse. Rotation and magnetic pressure may temporarily delay this process, allowing additional energy accumulation.

When critical conditions are reached, the core undergoes pair-instability collapse and explosive mass ejection. This event terminates the stable triple phase and initiates hypercritical accretion onto the neutron star, destabilization of the white dwarf, and the subsequent formation of a Kerr black hole and a new neutron star. The timing and energetics of the GRB are therefore directly linked to the preceding long-term evolution of the system.

\section{Evolution of the Three Body System}
\label{evolution}

In this section, we investigate the dynamical evolution of the three-body system introduced previously. The configuration consists of a compact, co-rotating inner binary maintained in a quasi-stationary Roche-lobe configuration, and a third component orbiting at larger separation on an eccentric trajectory. A rigorous analysis of the long-term dynamical stability of this hierarchical triple will be presented in a forthcoming study.

For the purposes of the present work, we proceed under the assumption that the hierarchical architecture remains dynamically stable over timescales sufficiently long to accommodate the secular evolution of the inner binary. In particular, we assume that the evolution of the inner orbit unfolds while the tertiary component—identified here as the accompanying neutron star—continues its orbital motion about the inner pair without inducing disruptive instabilities.

\begin{figure*}
  \centering
  \includegraphics[width=0.75\hsize,clip]{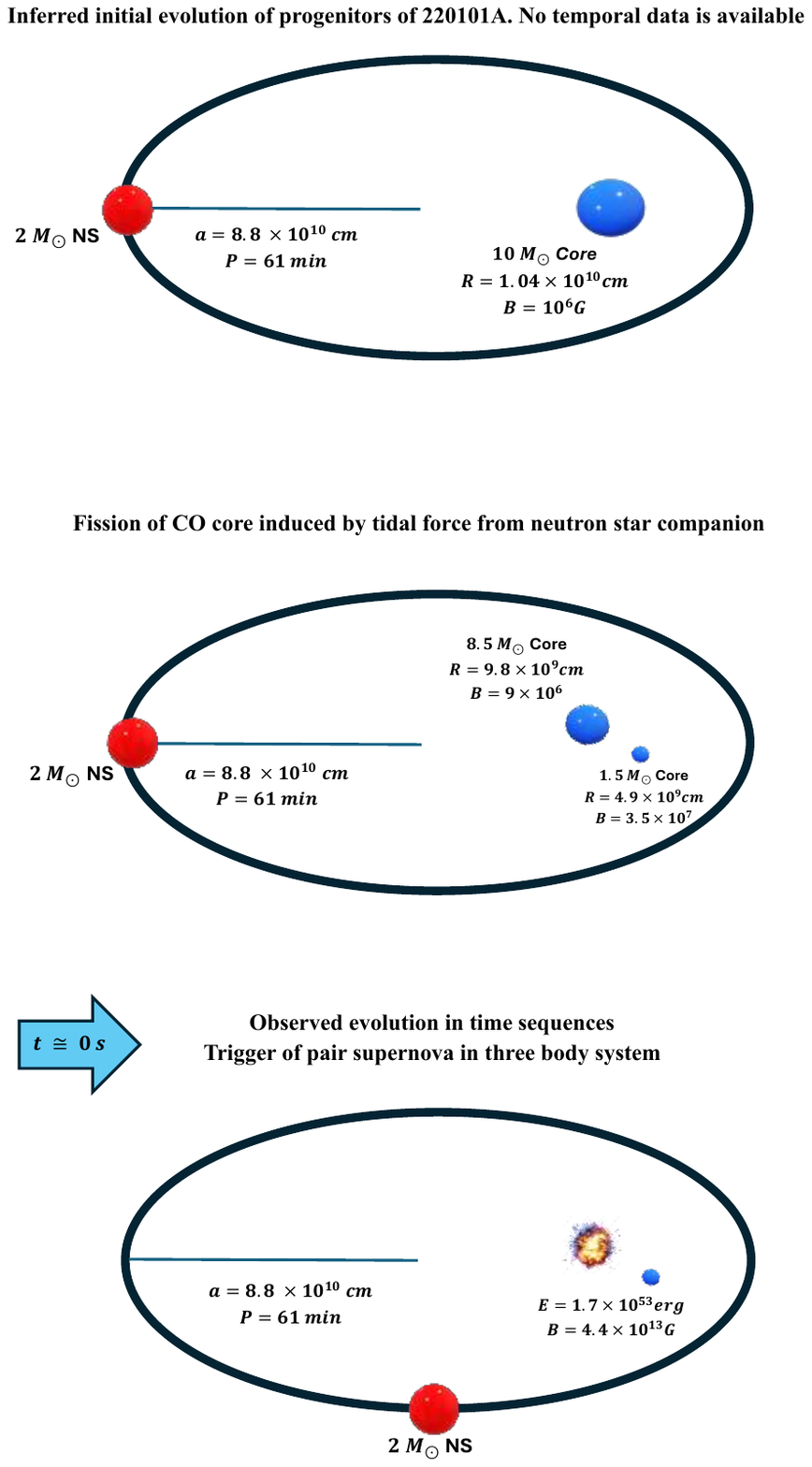}
 \caption{ The evolution of the binary system: from top to bottom, first,  The 10 $M_\odot$ core and 2 $M_\odot$ neutron star are initially at an eccentric orbit. Second, tidal interaction between the CO core and NS companion transfer orbital angular momentum from the orbit to CO core up to the point that the CO core reaches to bifurcation eccentricity and trigger splitting. The results is a three-body system composed of a new binary system with a 8.5 $M_\odot$ and 1.5 $M_\odot$ components. Third, the 8.5 $M_\odot$ component collapse and reaches to critical field and explode as a pair-SN.}\label{fig:orbit1}
\end{figure*}

\begin{figure*}[!tbp]
  \centering 
  \includegraphics[width=0.85\hsize,clip]{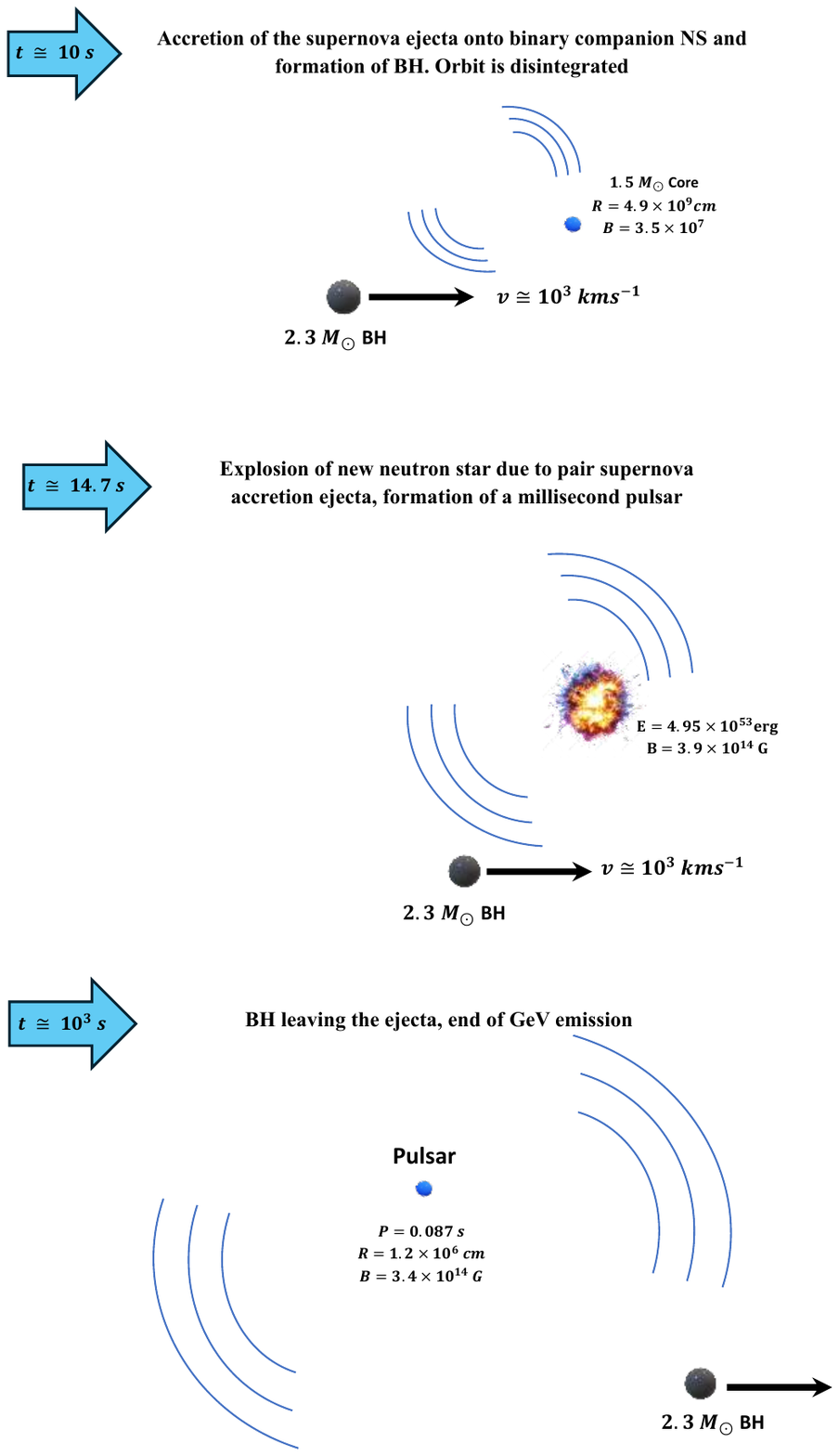}
  
  
\caption{The evolution of the binary system following the pair-instability supernova (pair-SN) explosion: Top:The pair-supernova leaves no compact remnant. The large ejecta mass disrupts the binary, ejecting the black hole with a velocity of order $10^3 \ kms^{-1}$.
Middle: Accretion of pair-SN ejecta onto the $1.5,M_\odot$ companion drives it unstable, triggering a secondary supernova explosion. Bottom: The black hole exits the ejecta, and the GeV emission ceases. The remnant of the $1.5,M_\odot$ supernova is a rapidly rotating millisecond pulsar.
}\label{fig:orbit2}
\end{figure*}

\subsection{The Pair-supernova}

The catastrophic evolution of the three-body system is initiated by the onset of instability in the more massive stellar component, whose core—of order $8$–$12 \ M_\odot$—undergoes gravitational collapse. Concurrent amplification of the internal magnetic field can drive the system toward a critical regime that may favor a magnetically mediated pair-instability supernova (PISN). In such an event, a thermonuclear runaway disrupts the star entirely, leaving no compact remnant (neither a black hole nor a neutron star). The total stellar mass is expelled, with an energy release approaching $10^{53} \ \mathrm{erg}$, predominantly in the form of photons and neutrinos. While classical PISN models are typically associated with helium core masses in the range $60$–$150 \ M_\odot$, it is conceivable that, under specific conditions—such as rapid rotation or strong magnetic-field amplification—analogous behavior may be triggered in lower-mass cores.

The initial magnetic field of the CO core can be as high as $10^6$. Assuming conservation of magnetic flux one can calculate magnetic field for both components of inner binary system. Collapse of CO core compress the magnetic field lines and intensify field strength. By manipulating flux conservation we can estimate the magnetic field during the collapse

\begin{equation}
\label{m1}
    B_p = \left( \frac{R_i}{R_p} \right)^2 B_i
\end{equation}
where $R_i$ and $B_i$ are initial value of radius and magnetic field of CO core and $R_p$ and $B_p$ are radius and magnetic field during the collapse. during the collapse intensified magnetic field radiate electromagnetic energy by consuming rotational energy.  Assuming the radiated energy is driven by dipole and quadrupole magnetic field the electro magnetic luminosity is 

\begin{align}
\label{Lum}
L_{\text{p}} &= \frac{dE_{\text{rot}}}{dt} = -I \Omega \dot{\Omega} \\
&= \frac{2}{3c^3} \Omega^4 B_{\text{dip}}^2 R_{\text{p}}^6 \sin^2 \chi_1 \left(1 + \eta^2 \frac{16}{45} \frac{R_{\text{p}}^2 \Omega^2}{c^2} \right),
\end{align}
where
\begin{equation}
\eta^2 = (\cos^2 \chi_2 + 10\sin^2 \chi_2) \frac{B_{\text{quad}}^2}{B_{\text{dip}}^2}.
\end{equation}
where $L_{p}$ denotes the electromagnetic luminosity that is equivalent to rate of rotational energy extraction from collapsing core. Here $B_{\text{dip}}$  and $B_{\text{quad}}$ are dipole and  quadrupole magnetic field respectively. $R_{\text{p}}$ is the instant radius of the core. The angles $\chi_1$ and $\chi_2$ indicate the inclinations of the magnetic moment components. The parameter $\eta$ quantifies the relative contribution of the quadrupole field to the total magnetic field, we refer to  \citet{2019ApJ...874...39W,2022ApJ...936..190W,2023ApJ...945...95W, 2022PhRvD.106h3002B} and the references therein. 

Recalling the definition of critical electric and magnetic fields $$ E_c =\frac{m_e^2 c^3}{e\hbar}, \ \  B_c = \frac{m_e^2 c^2}{e\hbar} = 4.4 \times 10^{13} \  G  $$ where $m_e$ and $e$ are the electron mass and charge respectively. This is the minimum value of the electromagnetic fields where can initiates vacuum polarization and pair production. To check that if the process of collapse can intensify  electromagnetic fields up to critical value we assumed a free fall profile for collapse and by solving  both equation \ref{m1} and \ref{Lum}, we look for a solution where the growing magnetic field which is inversely proportional to radius of collapsing CO core can reach to the critical field before crossing its black hole horizon. If such solution exist the optically thick plasma of $e^+,e^-$  accelerates in the strong induced electric field and explodes the newly formed core as a pair-supernova.

\begin{figure*}
\centering
\includegraphics[angle=0, scale=0.7]{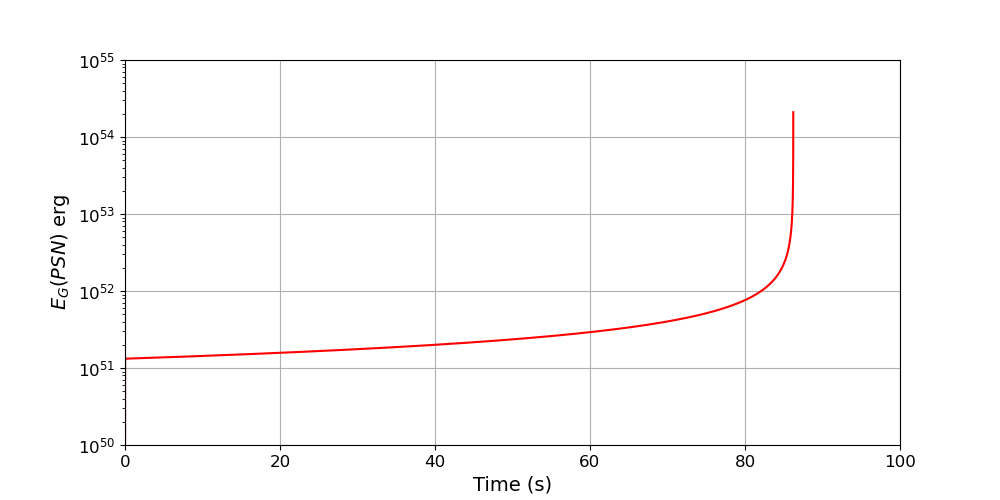}
\caption{ The gravitational energy released from a magnetized CO core before it reached to critical magnetic field }\label{fig:E_PSN02}
\end{figure*}

Figure \ref{fig:E_PSN02} illustrates the gravitational binding energy released by the CO core during collapse. To assess the onset of a pair-instability supernova, we track the temporal evolution of the magnetic field and compare it to the critical threshold $B_c$ prior to horizon formation.

For slowly rotating cores, the initial configuration is relatively compact, and the collapse does not amplify the magnetic field sufficiently to reach the critical regime required for a supernova explosion. Consequently, the core is expected to cross its event horizon before attaining $B_c$, precluding the occurrence of a pair-instability supernova.

In contrast, for rapidly rotating cores (with initial rotational periods shorter than $\sim 100$ min), the collapse can amplify the magnetic field to the critical level, triggering a pair-instability explosion as the core contracts to rotational periods of the order of a few seconds. In the case of anisotropic explosions, this characteristic timescale may be imprinted onto the supernova ejecta and subsequent accretion onto the binary companion, potentially giving rise to quasi-periodic modulations in relativistic jets, which could be observable in the MeV energy band.

\subsection{The Second Supernova}

Following this explosion a significant amount of relativistic ejecta spread all over the binary system and provides the less massive component of the binary with an intense phase of accretion. Soon after the accretion started the white dwarf (less massive component in inner binary system)  reach its critical mass and collapse and make the so called second supernova which we believe that is as strong as the pair-supernova. The velocity  of supernova ejecta has a stratified profile with low density outer layers has higher velocity $>0.1 \ c$ with respect to high density inner layers which expellee the ejecta with lower velocities $< 0.01 \ c$ \citep{1996snih.book.....A}. This supernova leave behind a compact object, a neutron star which accordingly develop a very strong magnetic field due to flux conservation. We can apply the process of free fall collapse similar to pair-supernova here and compute the final period, magnetic field and rotational energy of the neutron star when it collapse to the radius of $R = 1.2 \times 10^6$ cm typical of a neutron star with mass of 1.5 $M_\odot$.

\begin{figure*}
\centering
\includegraphics[angle=0, scale=0.7]{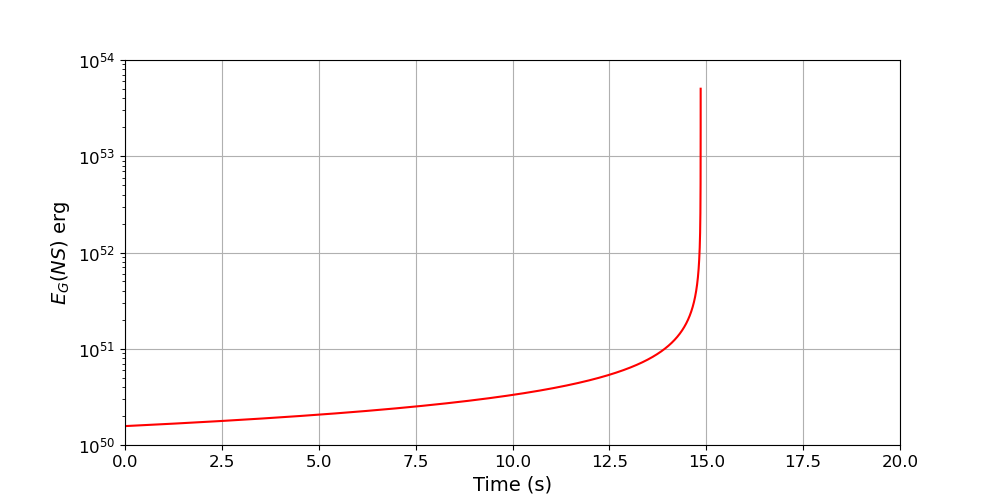}
\caption{ The gravitational energy released from a white dwarf  before it collapse to a neutron star }\label{fig:E_NS02}
\end{figure*}

Figure \ref{fig:E_NS02} represent gravitational energy available from white dwarf after it accrete enough material from pair supernova.  As we have seen in the pair-supernova initially fast core are small and can not develop high magnetic field and retain an object with nuclear density which resemble a normal pulsar. Slow rotating core can magnify their magnetic field up to $10^{16}$  which resembles a magnetar. These highly magnetize objects can make ultra relativistic jets and emits MeV photons.

\section{GRB 220101A: A clear and luminous manifestation of the model}

\begin{figure*}
\centering
\includegraphics[angle=0, scale=1.0]{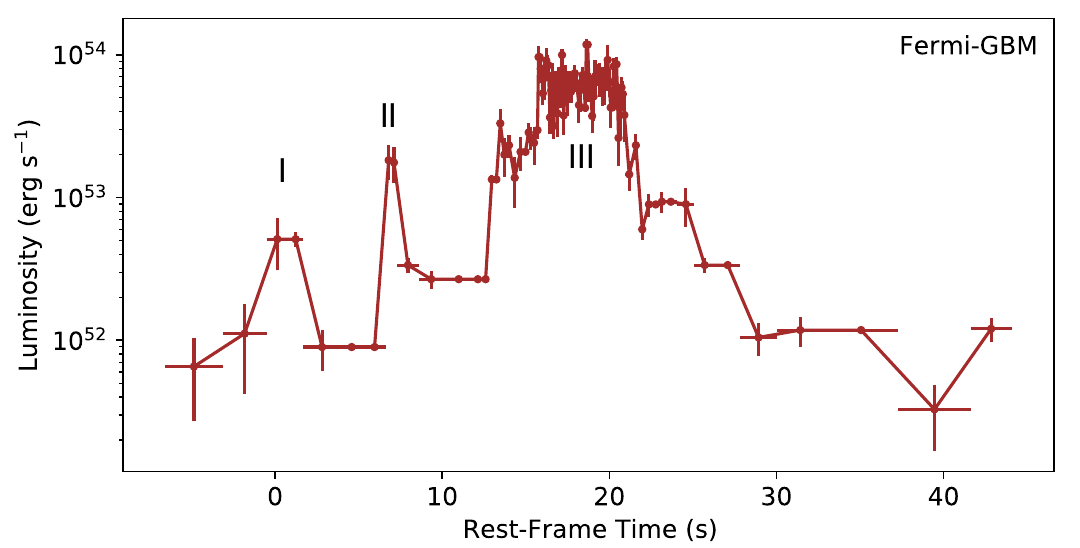}
\caption{ The Isotropic luminosity light-curve from Fermi-GBM of 1 keV to 10 MeV. Including the three episodes of I, II and III (Ruffini, et. al. in preparation}\label{fig:gbm}
\end{figure*}

After giving a general description of the model under consideration, in this section we applied this model to GRB 220101A which is one of the most energetic gamma-ray bursts recorded, located at a redshift of $z = 4.61$, with an isotropic energy of $E_{iso} = 4 × 10^{54}$ erg  \citep{2022GCN.31353....1F, 2022GCN.31360....1L}. The event is interpreted within the BdHN model based on a $10 \ M_\odot$ carbon-oxygen (CO) core and a $2 \ M_\odot$ neutron star (NS) companion \citep{2022GCN.31648....1R, 2022GCN.31465....1R}. Figure \ref{fig:gbm} illustrate MeV data observed by GBM detectors  on board Fermi satellite. Based on the model presented in this study we can identify three major events on this light curve. In the event I  the collapse of the CO core triggers the pair-supernova (Figure \ref{fig:gbm}). The CO core collapsing under its own gravity and reach conditions where the temperature and density trigger prolific electron-positron pair production. The entire stellar mass is ejected, releasing up to $10^{53} $ erg of energy.

After pair-supernova explosion the expelled ejecta reaches to the white dwarf in approximately 9 second and make it unstable under the gravitational collapse and trigger the birth of second supernova. This make the second peak in the figure \ref{fig:gbm} which we denotes by event II. 
We can estimate the ejection velocity of relativistic material being accreted with the less massive component 
\begin{equation}
v_{ej} = \frac{a}{\Delta t} = 2 \times 10^9 = 0.067c
\end{equation}
The event II leave behind a compact object, a neutron star which accordingly develop a very strong magnetic field due to flux conservation. 

\begin{figure*}
\centering
\includegraphics[angle=0, scale=0.6]{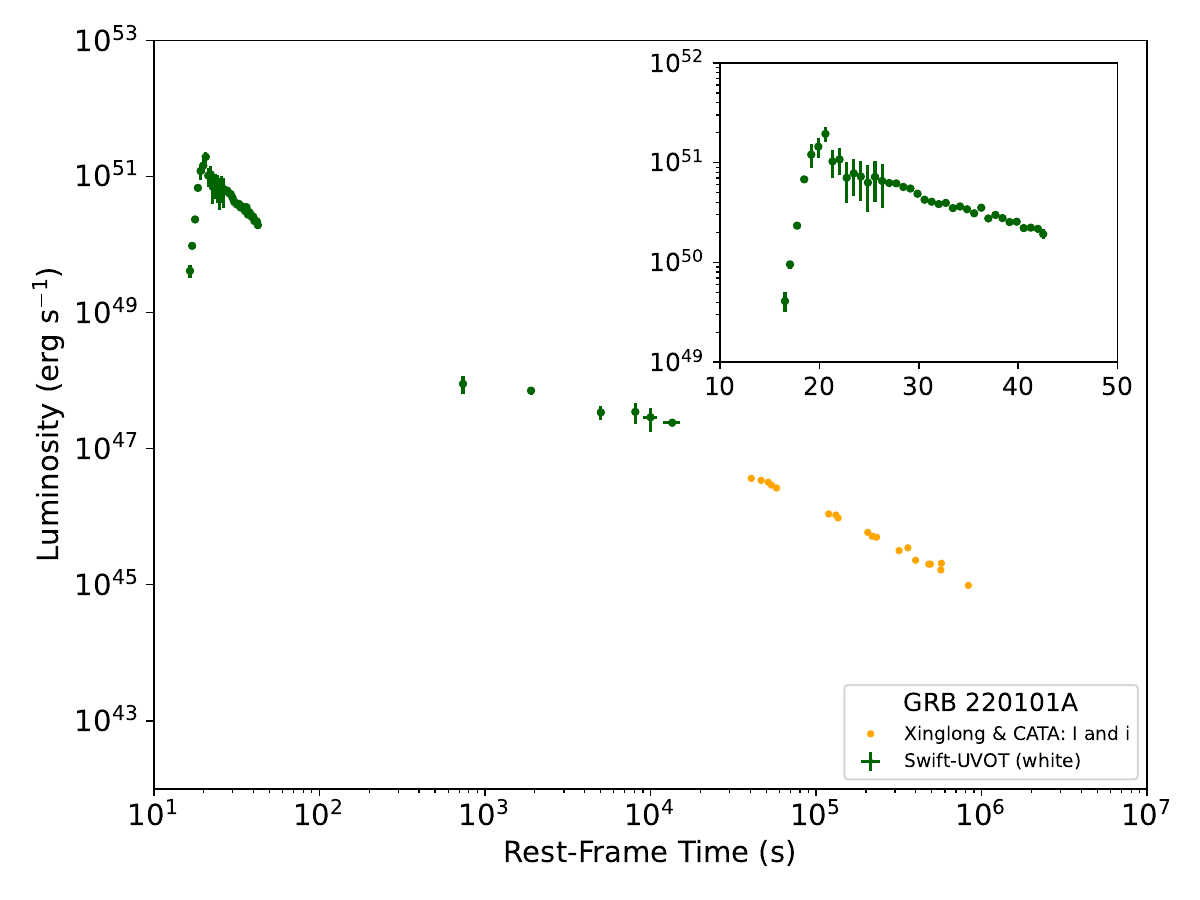}
\caption{The first 100 second of optical light curve of GRB 220101A taken by Swift-UVOT. During the raising part it is assumed the underlying NS absorb angular momentum from ejecta provided by supernova and spin up from 52 ms to 1 ms}
\label{fig:optical_220101A}
\end{figure*}

The optical light curve of GRB 220101A has been observed by Swift-UVOT from 16 up to $10^4$ second rest frame time (Figure \ref{fig:optical_220101A}).   With reference to event II in MeV light curve, the collapse process have had about 10 second to perform a compact core of the neutron star. During this collapse the NS core spin up with contraction and spin down with continues dipole and quadrupole radiation. Manipulating both equation \ref{m1} and \ref{Lum}, we estimate the angular velocity of the neutron star at the beginning of the optical observation by Swift-UVOT at the time around 16 second and have found that the neutron star is at its slowest state with a period of 87 ms. 

\begin{figure*}
\includegraphics[angle=0, scale=0.35]{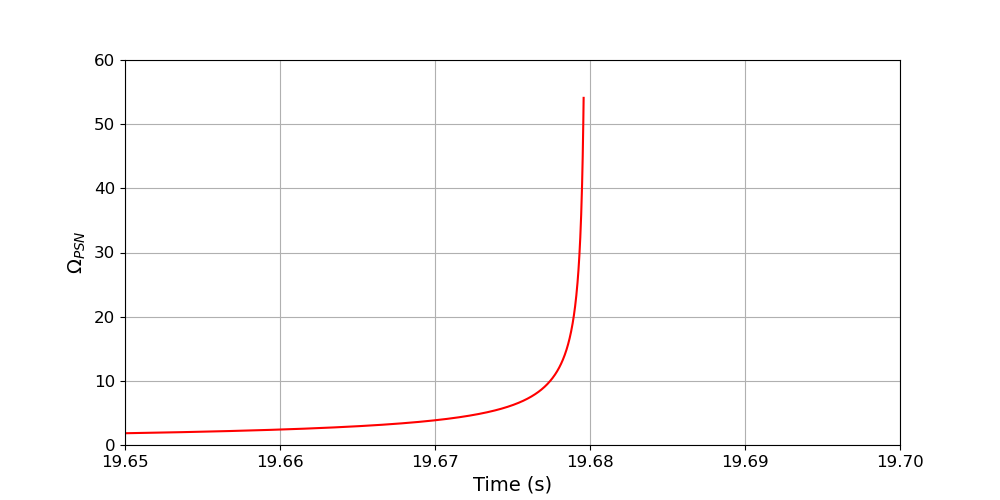}
\includegraphics[angle=0, scale=0.35]{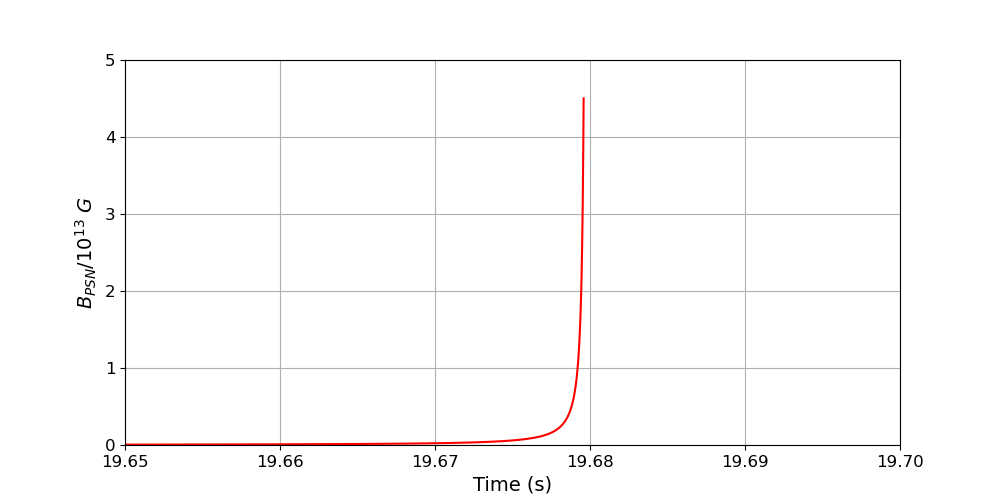}
\includegraphics[angle=0, scale=0.35]{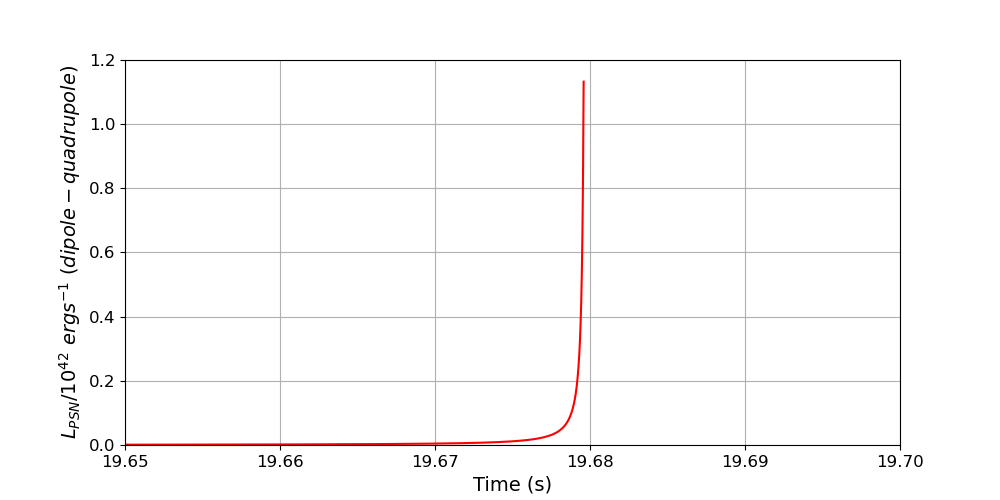}
\includegraphics[angle=0, scale=0.35]{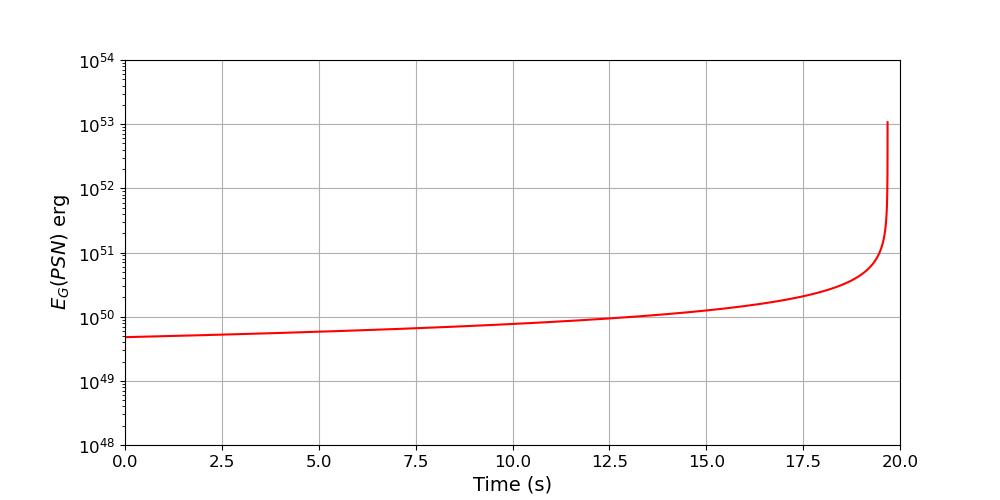}
\caption{Plots of angular velocity ($\Omega_{PSN}$), magnetic field ($B_{PSN}$), dipole luminosity ($L_{PSN}$) and released gravitational energy ($E_g$ PSN) of CO core during its collapse to a pair-SN vs. collapse time. }
\label{fig:specra1}
\end{figure*}

\begin{figure*}
\includegraphics[angle=0, scale=0.35]{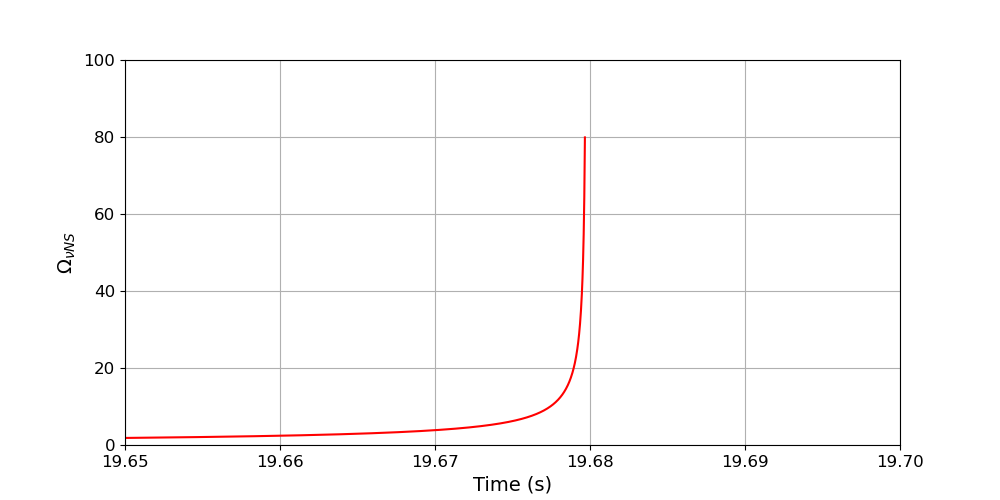}
\includegraphics[angle=0, scale=0.35]{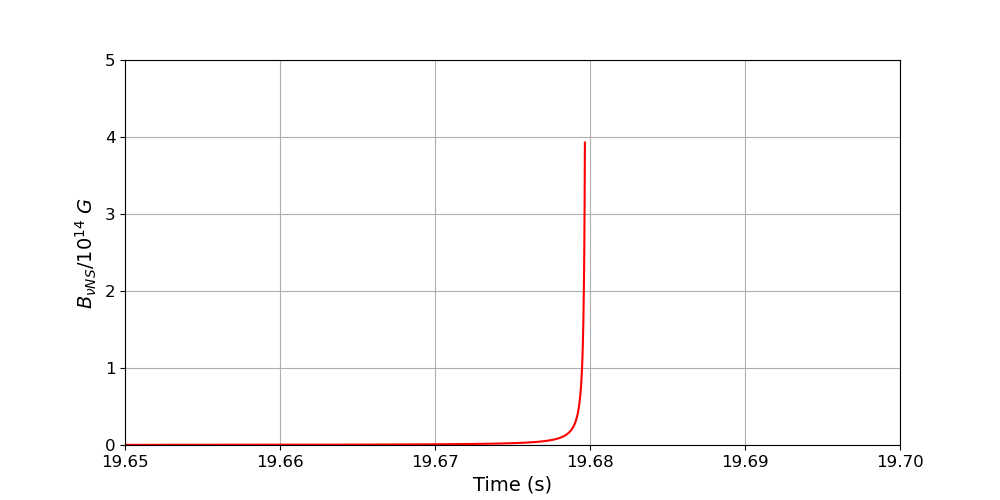}
\includegraphics[angle=0, scale=0.35]{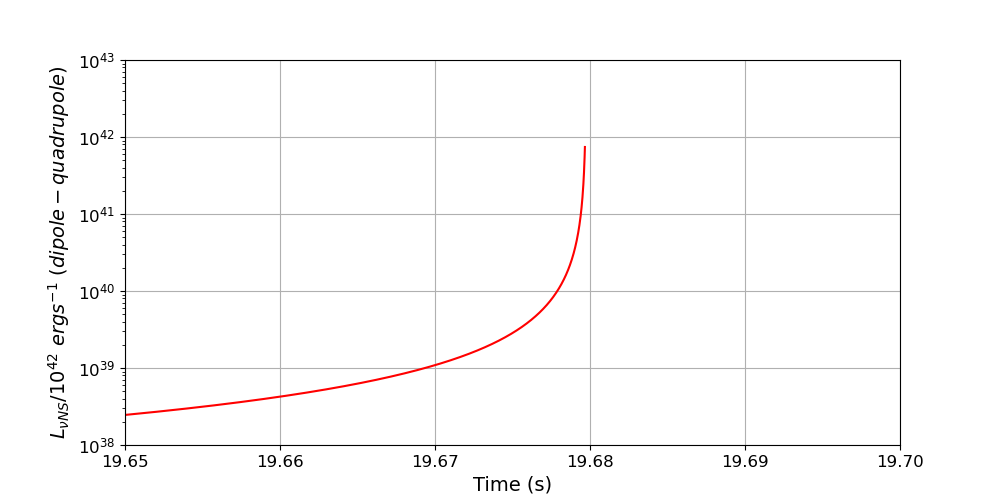}
\includegraphics[angle=0, scale=0.35]{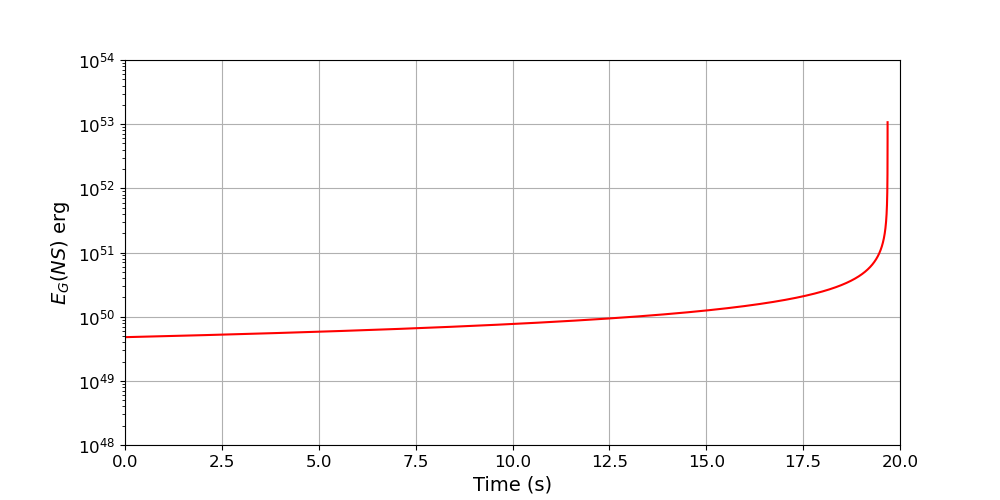}
\caption{Plots of angular velocity ($\Omega_{\nu NS}$), magnetic field ($B_{\nu NS}$), dipole luminosity ($L_{\nu NS}$) and released gravitational energy ($E_g$ ($\nu NS$) of white dwarf during its collapse to a neutron star vs. collapse time. }
\label{fig:specra2}
\end{figure*}

To reproduce the emission luminosity of $10^{49}$ erg/s with a magnetic dipole rotator with such short period of 0.052 second there should be at least a magnetic field with strength more than $10^{17}$ Gauss. During the raising part of light curve the neutron star undergoes rapid spin-up due to mass accretion from its surrounding environment. Over the course of 10 seconds, Accretion  dramatically  change its angular momentum. Assuming the neutron star at its fastest spinning at the maximum of the optical light curve accreted plasma interact with the magnetic field and reduce it down to $10^{13}$ G at maximum of optical light curve. During this period magnetic field is over critical ($B > B_c$). This overcritical field  and as mentioned before is the key factor to reproduce the ultra relativistic prompt mission (UPE), As figure \ref{fig:gbm} represent the period of overcritical field is coincide with the event III in Swift-GBM observation.

\section{Summary and Conclusion}

This manuscript presents a substantial extension of the Binary-Driven Hypernova (BdHN) paradigm by introducing a novel progenitor channel for the most energetic gamma-ray bursts (GRBs), specifically exemplified by GRB 220101A. The central idea is that a hierarchical triple system—composed of a magnetized carbon–oxygen (CO) core, a white dwarf (WD), and a neutron star (NS)—can naturally account for observational features that remain unexplained within conventional binary frameworks. This work integrates stellar evolution, relativistic astrophysics, and magnetohydrodynamics into a unified scenario characterized by multi-stage explosive phenomena and complex dynamical interactions.

At the core of the model lies the evolution of a compact triple system formed through standard channels of massive stellar evolution, including mass transfer, common-envelope evolution, and supernova-induced compact object formation. The resulting configuration — a CO core $\sim 8–12 \ M_\odot$, a WD companion, and a more distant NS—is shown to be dynamically stable over astrophysically relevant timescales, satisfying hierarchical stability criteria and surviving natal kicks under favorable conditions. 

The catastrophic phase is initiated by the onset of pair-instability in the magnetized CO core. As the core contracts and its temperature exceeds $\sim 10^9$ K, electron–positron pair production softens the equation of state, triggering a dynamical collapse. Crucially, emphasize the role of rotation and magnetic field amplification: for sufficiently rapid rotation, magnetic flux conservation can elevate the magnetic field to the quantum critical value $Bc \sim 4.4 × 10^{13} G$, enabling vacuum polarization and runaway pair creation. This leads to a magnetically mediated pair-instability supernova (pair-SN), which completely disrupts the core without leaving a compact remnant and releases energies approaching $10^{53}$ erg. 

The ejecta from this primary explosion interact with both companions, producing two distinct evolutionary branches. First, hypercritical accretion onto the neutron star companion drives it beyond its stability limit, inducing collapse into a Kerr black hole. This process powers the prompt high-energy (GeV) emission through relativistic outflows and magnetohydrodynamic energy extraction mechanisms. Second, accretion onto the white dwarf leads to its destabilization and subsequent collapse, triggering a secondary supernova. This event produces a newly formed neutron star $\nu$ NS, which rapidly spins up and develops an intense magnetic field via flux conservation, potentially reaching magnetar-level strengths.

A defining feature of this model is the temporal sequencing of these events. The secondary supernova is predicted to occur approximately $\sim 10–15$ seconds after the initial pair-SN, a timescale consistent with observed multi-peak structures in GRB light curves. The $\nu$ NS is identified as the dominant source of early optical emission, with its rotational energy and magnetic dipole radiation accounting for luminosities up to $\sim 10^{49} \  erg s^{-1}$. 

Another critical observational prediction concerns the absence of radioactive nickel decay signatures in the late-time optical light curve. Unlike conventional core-collapse supernovae, where $^{56}Ni$ decay governs the luminosity decline, the pair-SN in this model does not synthesize significant nickel. Consequently, the late-time emission is instead powered by pulsar activity, providing a distinctive diagnostic for identifying such events.

Application of the model to GRB 220101A demonstrates strong consistency with observational data, including its high isotropic energy $\sim 4 × 10^{54} erg $, rapid onset of high-energy emission, and anomalous optical behavior. The identification of three principal emission episodes—pair-SN, secondary SN, and pulsar-driven emission—offers a coherent interpretation of the multi-wavelength light curve.

\acknowledgments
MH gratefully acknowledges the full financial support by the Estonian Research Council through the Personal Research Funding project PRG2608 and the Center of Excellence TK202 "Foundations of the Universe".

\bibliography{fission_Refs}

\end{document}